\documentclass[twocolumn,showpacs,preprintnumbers,amsmath,amssymb]{revtex4}

\usepackage{graphicx}
\usepackage{dcolumn}
\usepackage{bm}

\begin{document}
\title{Coherent back-scattering near the two-dimensional metal-insulator transition}
\author{Maryam Rahimi, S. Anissimova, M.~R. Sakr, and S.~V. Kravchenko}
\affiliation{Physics Department, Northeastern University, Boston, Massachusetts 02115}
\author{T.~M. Klapwijk}
\affiliation{Department of Applied Physics, Delft University of Technology, 2628 CJ Delft, The Netherlands}
\date{\today}
\begin{abstract}
We have studied corrections to conductivity due to the coherent backscattering in low-disordered two-dimensional electron systems in silicon for a range of electron densities including the vicinity of the metal-insulator transition, where the dramatic increase of the spin susceptibility has been observed earlier.  We show that the corrections, which exist deeper in the metallic phase, weaken upon approaching to the transition and practically vanish at the critical density, thus suggesting that the localization is suppressed near and at the transition even in zero field.
\end{abstract}

\pacs{71.30.+h,73.20.Fz,73.20.Jc}
\maketitle

The metallic conductivity and the metal-insulator transition (MIT) in two-dimensional (2D) systems have been a subject of intense studies for almost a decade \cite{abrahams01}.  It is now well established that strong interactions between carriers are responsible for the metallic behavior, with quantitative agreement between the experiment and interaction-based theory \cite{zala01,punnoose02} reported for a variety of dilute 2D systems.  However, no quantitative theory is currently available in the most interesting regime: in the immediate vicinity of the transition, where electron-electron interactions cause strong changes in the system parameters including dramatic rise of the spin susceptibility \cite{shashkin01a,vitkalov02a,pudalov02a,zhu03} possibly leading to a spontaneous spin polarization at a finite electron density $n_\chi$ \cite{shashkin01a,vitkalov01}.  One of the open questions is the fate of the coherent backscattering (responsible for the weak localization) in the presence of strong interactions.  In some experiments on p-GaAs/AlGaAs heterostructures, negative magnetoresistance consistent with suppression of the coherent backscattering by a perpendicular magnetic field, $B_\perp$, has been reported at carrier densities as low as $4\times10^{10}$~cm$^{-2}$ corresponding to $r_s\approx10$, which has been interpreted as evidence of the insulating character of the ground state \cite{simmons00} (here $r_s\equiv E_c/E_F$ is the ratio between Coulomb and Fermi energies characterizing the strength of the interactions).  However, in ultra-clean p-GaAs/AlGaAs heterostructures with much lower carrier densities (down to $7\times10^{9}$~cm$^{-2}$), the negative magnetoresistance has been found to be negligible near the metal-insulator transition \cite{mills01} --- less than 3\% of its ``noninteracting'' value --- suggesting that the effects of the coherent backscattering are suppressed in this strongly-interacting system.

In this Letter, we report the first experimental studies of the weak-field magnetoconductance in high-mobility silicon metal-oxide-se\-mi\-con\-ductor field-effect transistors (MOSFETs) at temperatures down to about 40~mK and at electron densities down to the critical density for the MIT, $n_c=0.82\times10^{11}$~cm$^{-2}$.  In this regime, spin susceptibility sharply increases (and, possibly, even diverges at $n_\chi~\approx0.8\times10^{11}$~cm$^{-2}$ \cite{shashkin01a,vitkalov02a,vitkalov01}) indicating that properties of the system are dramatically altered by the interactions.  Previous studies of the coherent backscattering in this system were restricted to electron densities $n_s>1.5\times10^{11}$~cm$^{-2}$ \cite{brunthaler01}, which are outside of this interesting regime.  We have found that relatively far from the transition, the corrections to the conductivity due to the coherent backscattering are of their ``normal'' (non-interacting) strength, in agreement with earlier reports \cite{brunthaler01}.  However, once the transition is approached ($\delta\equiv n_s/n_c-1\lesssim0.1$), the corrections become progressively weaker, and in the immediate vicinity of the critical density, they practically vanish.  This suggests that the weak localization is already suppressed near and at the transition even in the absence of the perpendicular magnetic field, in agreement with recent data on ultra-high mobility p-GaAs/AlGaAs heterostructures \cite{mills01}.  The absence of localization at and just above the critical density may account for the existence of a flat separatrix between metallic and insulating states observed at $n_s=n_c$ in Refs.\cite{sarachik99,kravchenko00}.

We used three high-mobility samples from different wafers which all displayed similar behavior.  The data shown below were obtained on two samples (NH50.3064.9N and NH50.4064.7N) from different wafers; peak electron mobilities in both samples were close to $3.3\times10^4$~cm$^2$/Vs (at $T=0.1$~K).  The resistance was measured by a standard four-terminal technique at very low frequencies (0.2 to 0.5~Hz) to minimize out-of-phase signal.  The voltage between potential probes was amplified by a factor of $10^3$ by a ultra low-noise battery-operated SR-560 differential amplifier with a symmetric input and then measured by a SR-830 lock-in detector.  There are two main experimental difficulties associated with low-temperature transport measurements: the first is to minimize the contact resistance, which tends to grow very high at mK temperatures, and the second is to ensure that the carriers are not overheated relative to the bath temperature.  The first problem was solved \cite{heemskerk98} by introducing gaps in the gate metalization which allowed for maintaining high electron density near the contacts even when the density in the main part of the sample was low.  The gaps were narrow enough ($<100$~nm) for the given gate-oxide thickness to provide a smoothly descending electrostatic potential from the high-density part to the low-density part.  To minimize the second problem, we filtered out the external noise by using two sets of $\pi$ filters.  The electron gas was cooled by 16 copper wires approximately 0.5~mm in diameter thermally coupled to the mixing chamber of a Kelvinox-100 dilution refrigerator.  To verify that the electrons were not overheated relative to the bath temperature, we studied the temperature dependence of the amplitude of the Shubnikov-de~Haas oscillations.  In Fig.~1, we show $\rho(B_\perp)$ dependences recorded at ``high'' electron density, $n_s=7.3\cdot10^{11}$~cm$^{-2}$~(a), and at ``low'' density, $n_s=1.17\cdot10^{11}$~cm$^{-2}$~(b).  We have carefully checked that the data were taken in the linear regime of response.  The temperature dependence of the normalized amplitude of the oscillations, $A/A_0$, is shown in the insets together with the theoretical dependences calculated using Lifshitz-Kosevich formula \cite{isihara86} (here $A_0=4\exp(-2\pi^2k_BT_D/\hbar\omega_c)$, $\omega_c=eB_\perp/mc$ is the cyclotron frequency, $m$ is the effective mass, and $T_D$ is the Dingle temperature).  As seen from the figures, the theoretical dependences coincide with the experimental data at both high and low electron densities down to the lowest accessed temperature $\approx40$~mK.  This confirms that the electrons were indeed ``cold'' in our experiments.  The power generated in the samples was maintained at about $10^{-15}$~W at low electron densities ($n_s\lesssim1.3\cdot10^{11}$~cm$^{-2}$) and at about $10^{-14}$~W at higher electron densities.  The measuring current ranged from 40~pA at the lowest densities to 2~nA at the highest ones.

\begin{figure}\vspace{-2.8cm}
\scalebox{0.35}{\includegraphics{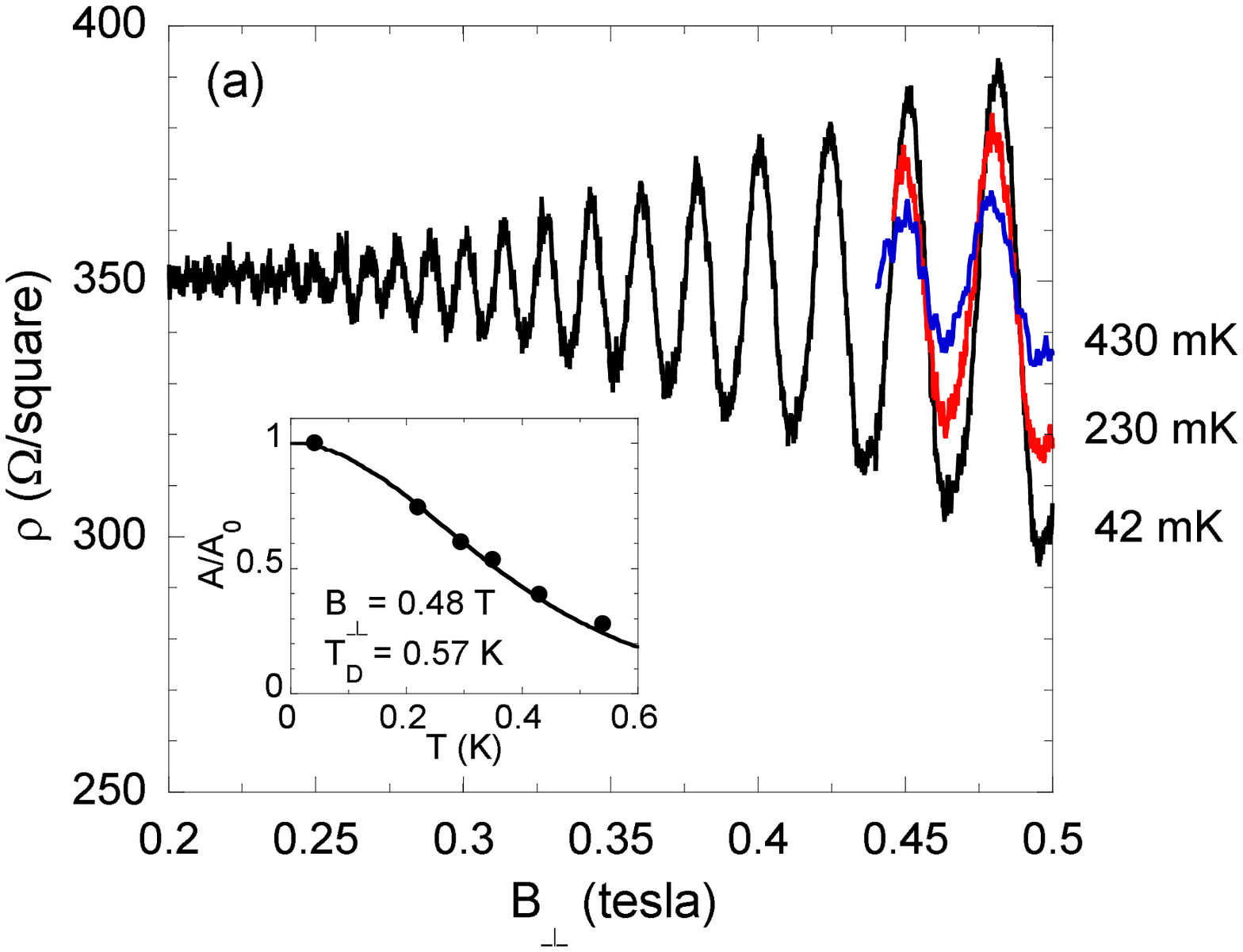}}\vspace{-4.2cm}
\scalebox{0.35}{\includegraphics{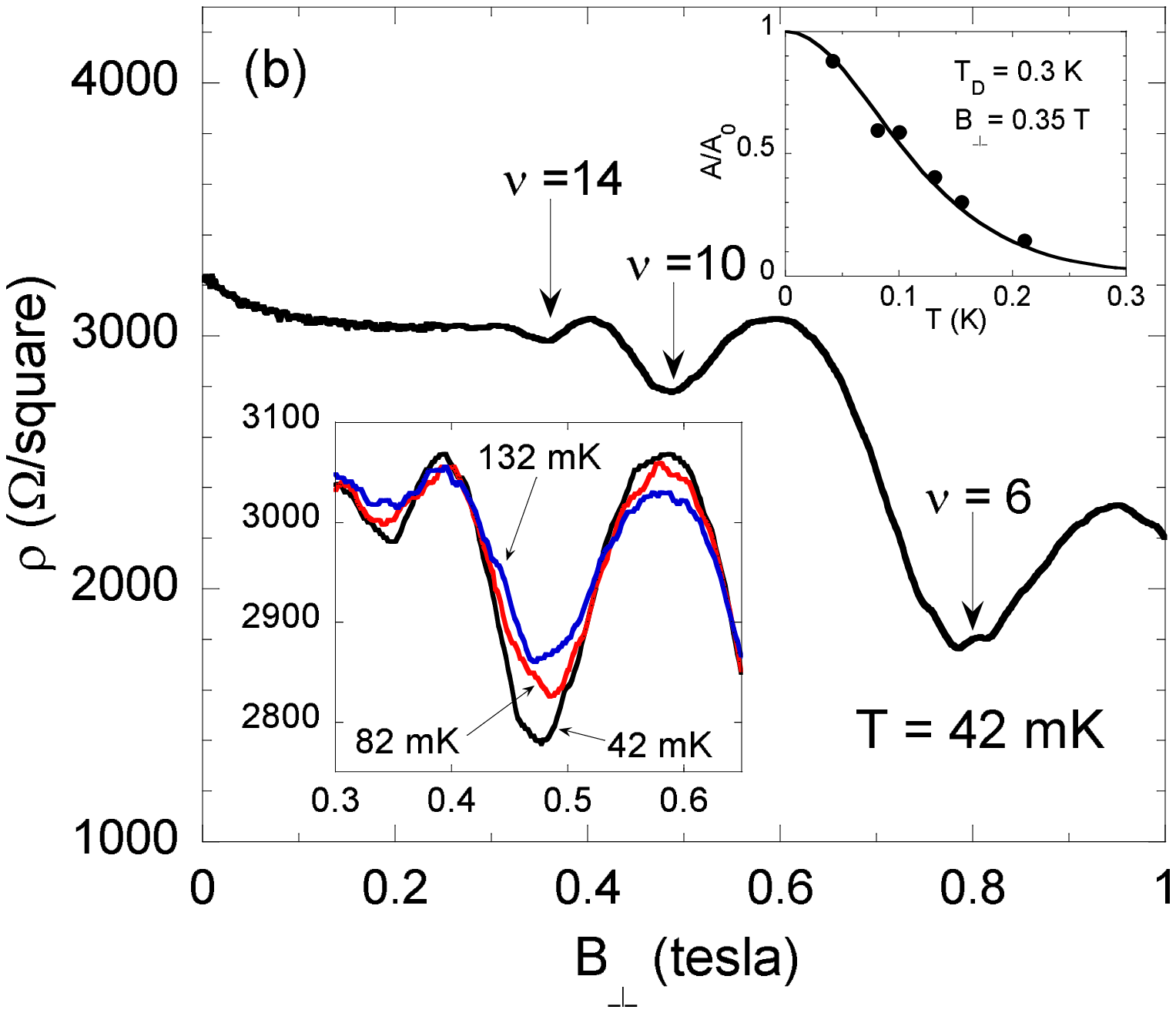}}\vspace{-2.2cm}
\caption{\label{fig.1.} Shubnikov-de~Haas oscillations in sample NH50.3064.9N at $n_s=7.3\cdot10^{11}$~cm$^{-2}$~(a) and $n_s=1.17\cdot10^{11}$~cm$^{-2}$~(b). The vertical arrows in (b) show the positions of the SdH minima corresponding to Landau filling factors $\nu=$~14, 10, and 6 as labeled.  Temperatures are indicated in the figures.  Measuring current is 2~nA~(a) and 200~pA~(b).  The experimentally measured amplitudes of the oscillations are shown in the insets by solid circles together with the calculated dependences (solid lines); the parameters used in calculations are indicated in the insets.  The lower inset in (b) shows a close-up view of two SdH oscillations corresponding to $\nu=$~14 and 10 measured at three temperatures.}
\end{figure}

In Fig.~2~(a), we show dependences of the longitudinal magnetoconductance, $\sigma$, on a perpendicular magnetic field recorded at $T=42$~mK for the range of electron densities on the metallic side of the metal-insulator transition.  All curves are symmetric about $B_\perp=0$.  At higher $n_s$ (upper curves), the characteristic magnetoconductance dip at zero magnetic field is seen.  It is associated with the effects of the coherent backscattering, or weak localization, which is present at $B_\perp=0$ but is suppressed by a weak perpendicular magnetic field.  Quantitatively similar $\sigma(B_\perp)$ dependences have been obtained on another sample, as shown in Fig.~2(b) (magnetoconductivity curves for this sample have been measured only at one direction of the magnetic field).

It is possible to fit the higher-$n_s$ magnetoconductance curves by the Hikami-Larkin formula \cite{hikami80}
\begin{equation}
\Delta\sigma=-\frac{e^2}{h}\;\frac{\alpha g_v}{\pi}\left[\Psi\left(\frac{1}{2}+\frac{\tau_B}{\tau}\right)
-\Psi\left(\frac{1}{2}+\frac{\tau_B}{\tau_\phi}\right)\right],\label{HL}
\end{equation}
where $\tau$ and $\tau_\phi$ are elastic and inelastic electron lifetimes, respectively, $\Psi$ is the Digamma function, $g_v$ is the valley degeneracy, $\tau_B=\hbar/4eB_\perp D$, and $D$ is the diffusion coefficient; the numeric coefficient $\alpha$ depends on the ratio of intravalley to intervalley scattering rates and is expected to be between 0.5 and 1.  An example of such fit is shown in Fig.~3~(a) (the top curve).  The value of the phase-breaking time used, $\tau_\phi=30$~ps, is consistent with experimental data for $\tau_\phi$ ($\sim20$ to 100~ps) obtained at similar temperatures on different 2D systems \cite{simmons00,brunthaler01,coleridge02}, although it is more than an order of magnitude shorter than that expected theoretically, $\tau_\phi\approx2\pi\hbar^2\sigma/e^2k_BT\,\text{ln}(\sigma h/2e^2)\sim10^{-9}$~s \cite{brunthaler01}.

As follows from Eq.~\ref{HL}, the magnitude of the dip is equal to $(e^2\alpha g_v/\pi h)$~ln$(\tau_\phi/\tau)$.  Using the above expression for the inelastic time, we obtain $\tau_\phi/\tau\propto1/\text{ln}(\sigma h/2e^2)$.  Therefore, the magnitude of the dip is expected to weakly (double-logarithmically) increase as the average conductivity decreases, provided the variations in electron density are small, as they are in our case.

This is not what is observed in the experiment: as one approaches the transition, the magnitude of the dip sharply drops, and at the critical electron density (the lowest curves in Figs.~2~(a,b)), the dip is no longer seen on the scale of these figures.  However, the shape of the magnetoconductivity does not change significantly with decreasing $n_s$.  This is illustrated by the middle curve in Fig.~3~(a) showing $\sigma(B_\perp)$ multiplied by six ($n_s=0.90\cdot10^{11}$~cm$^{-2}$), which makes it quantitatively similar to the upper curve.  This similarity demonstrates that the functional form of the $\sigma(B_\perp)$ dependence, described by the expression in brackets in Eq.~\ref{HL}, does not change noticeably as the density is reduced from $1.23\cdot10^{11}$ to $0.90\cdot10^{11}$~cm$^{-2}$; instead, it is the vertical scale of the effect that rapidly decreases upon approaching to the MIT.  At yet lower density, $n_s=0.82\cdot10^{11}$~cm$^{-2}$, the magnitude of the dip does not exceed 2\% of that for $n_s=1.23\cdot10^{11}$~cm$^{-2}$ (compare the upper and the lower curves in Fig.~3~(a)).  Therefore, in the immediate vicinity of the metal-insulator transition, the coherent backscattering in the metallic phase appears to be strongly suppressed even in zero magnetic field \cite{remark}.  This is in agreement with the results obtained in ultra-high-quality p-GaAs/AlGaAs heterostructure \cite{mills01}.

\begin{figure}\vspace{-1.07cm}
\scalebox{0.485}{\includegraphics{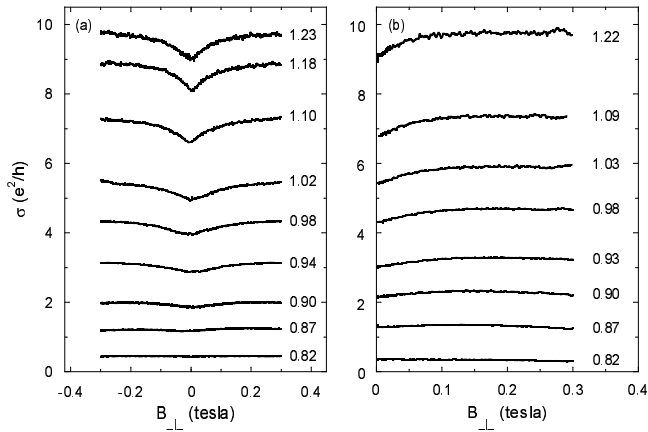}}\vspace{-6.4cm}
\caption{\label{fig.2.} Longitudinal magnetoconductivity in a weak perpendicular magnetic field at $T=42$~mK for a range of electron densities which are indicated near each curve in units of $10^{11}$~cm$^{-2}$ in sample NH50.3064.9N (a) and NH50.4064.7N (b).}
\end{figure}

In principle, suppression of the magnitude of the weak localization dip may be due to the inelastic scattering time being strongly reduced compared to that expected from the noninteracting theory.  However, our attempts to approximate magnetoconductivity close to the transition by the Hikami-Larkin formula with reduced $\tau_\phi$ have failed.  An example of such an attempt is shown in Fig.~3~(b).  The required magnitude of the dip can be achieved by choosing $\tau_\phi=6$~ps; however, the calculated dependence (the lower dashed curve) is much broader than the experimental one, and its curvature, $d^2\sigma/dB_\perp^2$, is positive in the entire range of magnetic fields, contrary to the experiment.  On the other hand, the experimental curve in Fig.~3~(b) can be fitted reasonably well (the upper dashed curve) by choosing the same parameters as for the highest-density curve in Fig.~3~(a), but with the prefactor ten times smaller ($\alpha=0.06$).  We therefore believe that it is the value of the prefactor, rather than $\tau_\phi$, which is responsible for the effect.

\begin{figure}\vspace{-2.53cm}
\scalebox{0.32}{\includegraphics{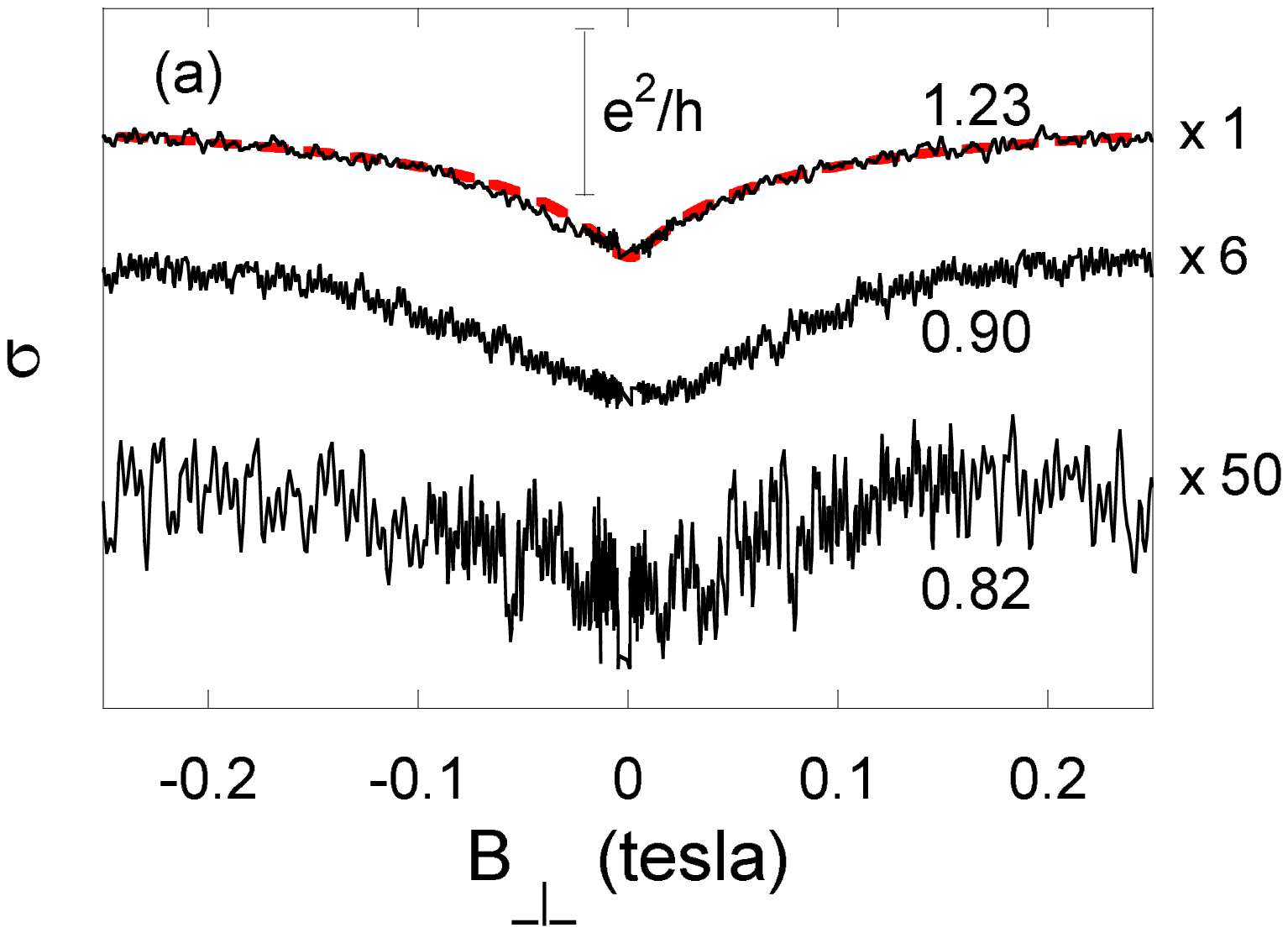}}\vspace{-5cm}
\scalebox{0.32}{\includegraphics{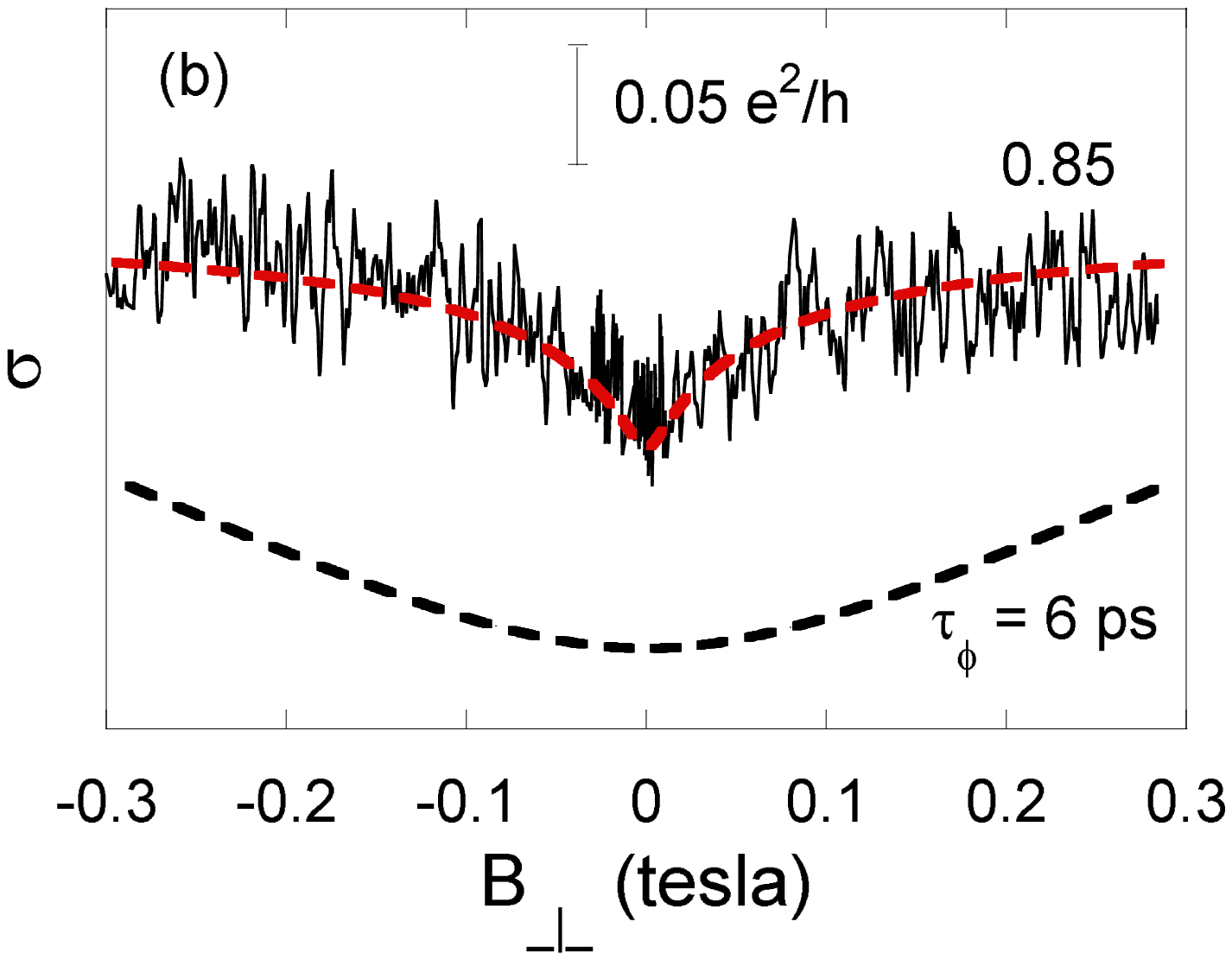}}\vspace{-2.5cm}
\caption{\label{fig.3.} Magnetoconductivity in sample NH50.3064.9N at $T=42$~mK and at different electron densities indicated near the curves in units of $10^{11}$~cm$^{-2}$.  (a)~The curves are vertically shifted for clarity, and two lower curves are multiplied by 6 (the middle one) and 50 (the lower one), which makes them quantitatively similar to the upper (unmodified) curve.  The thick dashed line is the fit by Eq.~\protect\ref{HL} with $\alpha=0.6$ and $\tau_\phi=30$~ps. (b)~Unsuccessful attempt to fit the experimental data for $n_s=0.85\cdot10^{11}$~cm$^{-2}$ by Eq.~\protect\ref{HL} with the same prefactor $\alpha=0.6$ as above but reduced $\tau_\phi=6$~ps (the lower dashed curve, which is shifted for clarity).  However, the data can be fitted reasonably well by choosing $\tau_\phi=30$~ps as in (a), but with the prefactor ten times smaller ($\alpha=0.06$; the upper dashed curve).}
\end{figure}

As we have already mentioned, experimental data on Si MOSFETs reveal a sharp increase of the spin susceptibility at $n_s\rightarrow n_\chi\approx0.8\cdot10^{11}$~cm$^{-2}$, {\it i.e.}, in the same region of densities where we observe the suppression of the weak localization.  Due to the strong renormalization of the effective mass \cite{shashkin02,shashkin03}, the interaction parameter $r_s$ reaches at least 50 near $n_\chi$ (which in our samples practically coincides with the critical density for the metal-insulator transition).  However, samples of a lower quality become strongly localized at densities noticeably higher than $n_\chi$, where the effective mass remains close to its band value and $r_s$ does not exceed 10.  Coherent backscattering in such samples is still present near the transition to a strongly localized state; there is no temperature-independent separatrix, and resistance of some of seemingly-metallic curves turns up at low temperatures \cite{prus02a}.  Similar behavior is observed in p-GaAs/AlGaAs heterostructures: ``ordinary'' weak localization corrections and low-temperature up-turns of the resistance have been reported in samples with relatively high transition density $\approx4.6\cdot10^{10}$~cm$^{-2}$ \cite{simmons00}, while no up-turn of the resistance and practically no weak localization corrections have been seen in higher-quality heterostructures with almost an order of magnitude lower transition density $\approx6\cdot10^9$~cm$^{-2}$ \cite{mills01}.

In summary, not too close to the transition, at $\delta\gtrsim0.1$, there are no principal deviations from the ``conventional'' magnetoconductance associated with the suppression of the coherent backscattering by a perpendicular magnetic field.  This fact, although known from earlier measurements \cite{simmons00,senz00,brunthaler01,coleridge02}, is not trivial, because the weak localization theory was developed for non- or weakly-interacting electron systems, while in our case, the energy of electron-electron interactions exceeds the Fermi energy by more than an order of magnitude.  However, in the immediate vicinity of the MIT, where the role of interactions is especially strong, the magnitude of the weak localization corrections sharply drops, and near the critical electron density, the corrections do not exceed 2\% of their ``noninteracting'' value.  This confirms similar results in an ultra-clean p-GaAs/AlGaAs heterostructure \cite{mills01} and thus establishes their universality.  The absence of the localization corrections at the metal-insulator transition may account for the existence of a flat separatrix between metallic and insulating states observed in Refs.\cite{sarachik99,kravchenko00}.

We are grateful to D. Heiman and A.~A. Shashkin for useful discussions and to D. Basiaga for technical assistance.  This work was supported by NSF grants DMR-9988283 and DMR-0129652 and the Sloan Foundation.


\begin{thebibliography}{21}
\expandafter\ifx\csname natexlab\endcsname\relax\def\natexlab#1{#1}\fi
\expandafter\ifx\csname bibnamefont\endcsname\relax
  \def\bibnamefont#1{#1}\fi
\expandafter\ifx\csname bibfnamefont\endcsname\relax
  \def\bibfnamefont#1{#1}\fi
\expandafter\ifx\csname citenamefont\endcsname\relax
  \def\citenamefont#1{#1}\fi
\expandafter\ifx\csname url\endcsname\relax
  \def\url#1{\texttt{#1}}\fi
\expandafter\ifx\csname urlprefix\endcsname\relax\def\urlprefix{URL }\fi
\providecommand{\bibinfo}[2]{#2}
\providecommand{\eprint}[2][]{\url{#2}}

\bibitem[{\citenamefont{Abrahams et~al.}(2001)\citenamefont{Abrahams,
  Kravchenko, and Sarachik}}]{abrahams01}
\bibinfo{author}{\bibfnamefont{E.}~\bibnamefont{Abrahams}},
  \bibinfo{author}{\bibfnamefont{S.~V.} \bibnamefont{Kravchenko}},
  \bibnamefont{and} \bibinfo{author}{\bibfnamefont{M.~P.}
  \bibnamefont{Sarachik}}, \bibinfo{journal}{Rev.\ Mod.\ Phys.}
  \textbf{\bibinfo{volume}{73}}, \bibinfo{pages}{251} (\bibinfo{year}{2001}).

\bibitem[{\citenamefont{Zala et~al.}(2001)\citenamefont{Zala, Narozhny, and
  Aleiner}}]{zala01}
\bibinfo{author}{\bibfnamefont{G.}~\bibnamefont{Zala}},
  \bibinfo{author}{\bibfnamefont{B.~N.} \bibnamefont{Narozhny}},
  \bibnamefont{and} \bibinfo{author}{\bibfnamefont{I.~L.}
  \bibnamefont{Aleiner}}, \bibinfo{journal}{Phys.\ Rev.\ B}
  \textbf{\bibinfo{volume}{64}}, \bibinfo{pages}{214204}
  (\bibinfo{year}{2001}).

\bibitem[{\citenamefont{Punnoose and Finkelstein}(2002)}]{punnoose02}
\bibinfo{author}{\bibfnamefont{A.}~\bibnamefont{Punnoose}} \bibnamefont{and}
  \bibinfo{author}{\bibfnamefont{A.~M.} \bibnamefont{Finkelstein}},
  \bibinfo{journal}{Phys.\ Rev.\ Lett.} \textbf{\bibinfo{volume}{88}},
  \bibinfo{pages}{016802} (\bibinfo{year}{2002}).

\bibitem[{\citenamefont{Shashkin et~al.}(2001)\citenamefont{Shashkin,
  Kravchenko, Dolgopolov, and Klapwijk}}]{shashkin01a}
\bibinfo{author}{\bibfnamefont{A.~A.} \bibnamefont{Shashkin}},
  \bibinfo{author}{\bibfnamefont{S.~V.} \bibnamefont{Kravchenko}},
  \bibinfo{author}{\bibfnamefont{V.~T.} \bibnamefont{Dolgopolov}},
  \bibnamefont{and} \bibinfo{author}{\bibfnamefont{T.~M.}
  \bibnamefont{Klapwijk}}, \bibinfo{journal}{Phys.\ Rev.\ Lett.}
  \textbf{\bibinfo{volume}{87}}, \bibinfo{pages}{086801}
  (\bibinfo{year}{2001}).

\bibitem[{\citenamefont{Vitkalov et~al.}(2002)\citenamefont{Vitkalov, Sarachik,
  and Klapwijk}}]{vitkalov02a}
\bibinfo{author}{\bibfnamefont{S.~A.} \bibnamefont{Vitkalov}},
  \bibinfo{author}{\bibfnamefont{M.~P.} \bibnamefont{Sarachik}},
  \bibnamefont{and} \bibinfo{author}{\bibfnamefont{T.~M.}
  \bibnamefont{Klapwijk}}, \bibinfo{journal}{Phys.\ Rev.\ B}
  \textbf{\bibinfo{volume}{65}}, \bibinfo{pages}{201106(R)}
  (\bibinfo{year}{2002}).

\bibitem[{\citenamefont{Pudalov et~al.}(2002)\citenamefont{Pudalov, Gershenson,
  Kojima, Butch, Dizhur, Brunthaler, Prinz, and Bauer}}]{pudalov02a}
\bibinfo{author}{\bibfnamefont{V.~M.} \bibnamefont{Pudalov}},
  \bibinfo{author}{\bibfnamefont{M.~E.} \bibnamefont{Gershenson}},
  \bibinfo{author}{\bibfnamefont{H.}~\bibnamefont{Kojima}},
  \bibinfo{author}{\bibfnamefont{N.}~\bibnamefont{Butch}},
  \bibinfo{author}{\bibfnamefont{E.~M.} \bibnamefont{Dizhur}},
  \bibinfo{author}{\bibfnamefont{G.}~\bibnamefont{Brunthaler}},
  \bibinfo{author}{\bibfnamefont{A.}~\bibnamefont{Prinz}}, \bibnamefont{and}
  \bibinfo{author}{\bibfnamefont{G.}~\bibnamefont{Bauer}},
  \bibinfo{journal}{Phys.\ Rev.\ Lett.} \textbf{\bibinfo{volume}{88}},
  \bibinfo{pages}{196404} (\bibinfo{year}{2002}).

\bibitem[{\citenamefont{Zhu et~al.}(2003)\citenamefont{Zhu, Stormer, Pfeiffer,
  Baldwin, and West}}]{zhu03}
\bibinfo{author}{\bibfnamefont{J.}~\bibnamefont{Zhu}},
  \bibinfo{author}{\bibfnamefont{H.~L.} \bibnamefont{Stormer}},
  \bibinfo{author}{\bibfnamefont{L.~N.} \bibnamefont{Pfeiffer}},
  \bibinfo{author}{\bibfnamefont{K.~W.} \bibnamefont{Baldwin}},
  \bibnamefont{and} \bibinfo{author}{\bibfnamefont{K.~W.} \bibnamefont{West}},
  \bibinfo{journal}{Phys.\ Rev.\ Lett.} \textbf{\bibinfo{volume}{90}},
  \bibinfo{pages}{056805} (\bibinfo{year}{2003}).

\bibitem[{\citenamefont{Vitkalov et~al.}(2001)\citenamefont{Vitkalov, Zheng,
  Mertes, Sarachik, and Klapwijk}}]{vitkalov01}
\bibinfo{author}{\bibfnamefont{S.~A.} \bibnamefont{Vitkalov}},
  \bibinfo{author}{\bibfnamefont{H.}~\bibnamefont{Zheng}},
  \bibinfo{author}{\bibfnamefont{K.~M.} \bibnamefont{Mertes}},
  \bibinfo{author}{\bibfnamefont{M.~P.} \bibnamefont{Sarachik}},
  \bibnamefont{and} \bibinfo{author}{\bibfnamefont{T.~M.}
  \bibnamefont{Klapwijk}}, \bibinfo{journal}{Phys.\ Rev.\ Lett.}
  \textbf{\bibinfo{volume}{87}}, \bibinfo{pages}{086401}
  (\bibinfo{year}{2001}).

\bibitem[{\citenamefont{Simmons et~al.}(2000)\citenamefont{Simmons, Hamilton,
  Pepper, Linfield, Rose, and Ritchie}}]{simmons00}
\bibinfo{author}{\bibfnamefont{M.~Y.} \bibnamefont{Simmons}},
  \bibinfo{author}{\bibfnamefont{A.~R.} \bibnamefont{Hamilton}},
  \bibinfo{author}{\bibfnamefont{M.}~\bibnamefont{Pepper}},
  \bibinfo{author}{\bibfnamefont{E.~H.} \bibnamefont{Linfield}},
  \bibinfo{author}{\bibfnamefont{P.~D.} \bibnamefont{Rose}}, \bibnamefont{and}
  \bibinfo{author}{\bibfnamefont{D.~A.} \bibnamefont{Ritchie}},
  \bibinfo{journal}{Phys.\ Rev.\ Lett.} \textbf{\bibinfo{volume}{84}},
  \bibinfo{pages}{2489} (\bibinfo{year}{2000}).

\bibitem[{\citenamefont{{Mills, Jr.} et~al.}(2001)\citenamefont{{Mills, Jr.},
  Ramirez, Gao, Pfeiffer, West, and Simon}}]{mills01}
\bibinfo{author}{\bibfnamefont{A.~P.} \bibnamefont{{Mills, Jr.}}},
  \bibinfo{author}{\bibfnamefont{A.~P.} \bibnamefont{Ramirez}},
  \bibinfo{author}{\bibfnamefont{X.~P.~A.} \bibnamefont{Gao}},
  \bibinfo{author}{\bibfnamefont{L.~N.} \bibnamefont{Pfeiffer}},
  \bibinfo{author}{\bibfnamefont{K.~W.} \bibnamefont{West}}, \bibnamefont{and}
  \bibinfo{author}{\bibfnamefont{S.~H.} \bibnamefont{Simon}}
  (\bibinfo{year}{2001}), \bibinfo{note}{cond-mat/0101020}.

\bibitem[{\citenamefont{Brunthaler et~al.}(2001)\citenamefont{Brunthaler,
  Prinz, Bauer, and Pudalov}}]{brunthaler01}
\bibinfo{author}{\bibfnamefont{G.}~\bibnamefont{Brunthaler}},
  \bibinfo{author}{\bibfnamefont{A.}~\bibnamefont{Prinz}},
  \bibinfo{author}{\bibfnamefont{G.}~\bibnamefont{Bauer}}, \bibnamefont{and}
  \bibinfo{author}{\bibfnamefont{V.~M.} \bibnamefont{Pudalov}},
  \bibinfo{journal}{Phys.\ Rev.\ Lett.} \textbf{\bibinfo{volume}{87}},
  \bibinfo{pages}{096802} (\bibinfo{year}{2001}).

\bibitem[{\citenamefont{Sarachik and Kravchenko}(1999)}]{sarachik99}
\bibinfo{author}{\bibfnamefont{M.~P.} \bibnamefont{Sarachik}} \bibnamefont{and}
  \bibinfo{author}{\bibfnamefont{S.~V.} \bibnamefont{Kravchenko}},
  \bibinfo{journal}{Proc.\ Natl.\ Acad.\ Sci.\ USA}
  \textbf{\bibinfo{volume}{96}}, \bibinfo{pages}{5900} (\bibinfo{year}{1999}).

\bibitem[{\citenamefont{Kravchenko and Klapwijk}(2000)}]{kravchenko00}
\bibinfo{author}{\bibfnamefont{S.~V.} \bibnamefont{Kravchenko}}
  \bibnamefont{and} \bibinfo{author}{\bibfnamefont{T.~M.}
  \bibnamefont{Klapwijk}}, \bibinfo{journal}{Phys.\ Rev.\ Lett.}
  \textbf{\bibinfo{volume}{84}}, \bibinfo{pages}{2909} (\bibinfo{year}{2000}).

\bibitem[{\citenamefont{Heemskerk and Klapwijk}(1998)}]{heemskerk98}
\bibinfo{author}{\bibfnamefont{R.}~\bibnamefont{Heemskerk}} \bibnamefont{and}
  \bibinfo{author}{\bibfnamefont{T.~M.} \bibnamefont{Klapwijk}},
  \bibinfo{journal}{Phys.\ Rev.\ B} \textbf{\bibinfo{volume}{58}},
  \bibinfo{pages}{R1754} (\bibinfo{year}{1998}).

\bibitem[{\citenamefont{Isihara and Smr\v{c}ka}(1986)}]{isihara86}
\bibinfo{author}{\bibfnamefont{A.}~\bibnamefont{Isihara}} \bibnamefont{and}
  \bibinfo{author}{\bibfnamefont{L.}~\bibnamefont{Smr\v{c}ka}},
  \bibinfo{journal}{J.\ Phys.\ C} \textbf{\bibinfo{volume}{19}},
  \bibinfo{pages}{6777} (\bibinfo{year}{1986}).

\bibitem[{\citenamefont{Hikami et~al.}(1980)\citenamefont{Hikami, Larkin, and
  Nagaoka}}]{hikami80}
\bibinfo{author}{\bibfnamefont{S.}~\bibnamefont{Hikami}},
  \bibinfo{author}{\bibfnamefont{A.}~\bibnamefont{Larkin}}, \bibnamefont{and}
  \bibinfo{author}{\bibfnamefont{Y.}~\bibnamefont{Nagaoka}},
  \bibinfo{journal}{Prog.\ Theor.\ Phys.} \textbf{\bibinfo{volume}{63}},
  \bibinfo{pages}{707} (\bibinfo{year}{1980}).

\bibitem[{\citenamefont{Coleridge et~al.}(2002)\citenamefont{Coleridge,
  Sachrajda, and Zawadzki}}]{coleridge02}
\bibinfo{author}{\bibfnamefont{P.~T.} \bibnamefont{Coleridge}},
  \bibinfo{author}{\bibfnamefont{A.~S.} \bibnamefont{Sachrajda}},
  \bibnamefont{and} \bibinfo{author}{\bibfnamefont{P.}~\bibnamefont{Zawadzki}},
  \bibinfo{journal}{Phys.\ Rev.\ B} \textbf{\bibinfo{volume}{65}},
  \bibinfo{pages}{125328} (\bibinfo{year}{2002}).

\bibitem{remark} On the insulating side of the transition ($n_s<n_c$), there are also no signs of the weak-field positive magnetoconductance/negative magnetoresistance, which could be associated with the suppression of the coherent backscattering.  This regime was studied in detail by M. D'Iorio, V.~M. Pudalov, and S.~G. Semenchinsky, Phys.\ Lett.\ A {\bf 150}, 422 (1990); Phys.\ Rev.\ B\ {\bf 46}, 15992 (1992); S.~V. Kravchenko, J.~A.~A.~J. Perenboom, and V.~M. Pudalov, Phys.\ Rev.\ B\ {\bf 44}, 13513 (1991); V.~M. Pudalov, M. D'Iorio, and J.~W. Campbell, Surf.\ Sci.\ {\bf 305}, 107 (1994).

\bibitem[{\citenamefont{Shashkin et~al.}(2002)\citenamefont{Shashkin,
  Kravchenko, Dolgopolov, and Klapwijk}}]{shashkin02}
\bibinfo{author}{\bibfnamefont{A.~A.} \bibnamefont{Shashkin}},
  \bibinfo{author}{\bibfnamefont{S.~V.} \bibnamefont{Kravchenko}},
  \bibinfo{author}{\bibfnamefont{V.~T.} \bibnamefont{Dolgopolov}},
  \bibnamefont{and} \bibinfo{author}{\bibfnamefont{T.~M.}
  \bibnamefont{Klapwijk}}, \bibinfo{journal}{Phys.\ Rev.\ B}
  \textbf{\bibinfo{volume}{66}}, \bibinfo{pages}{073303}
  (\bibinfo{year}{2002}).

\bibitem[{\citenamefont{Shashkin et~al.}(2003)\citenamefont{Shashkin, Rahimi,
  Anissimova, Kravchenko, Dolgopolov, and Klapwijk}}]{shashkin03}
\bibinfo{author}{\bibfnamefont{A.~A.} \bibnamefont{Shashkin}},
  \bibinfo{author}{\bibfnamefont{M.}~\bibnamefont{Rahimi}},
  \bibinfo{author}{\bibfnamefont{S.}~\bibnamefont{Anissimova}},
  \bibinfo{author}{\bibfnamefont{S.~V.} \bibnamefont{Kravchenko}},
  \bibinfo{author}{\bibfnamefont{V.~T.} \bibnamefont{Dolgopolov}},
  \bibnamefont{and} \bibinfo{author}{\bibfnamefont{T.~M.}
  \bibnamefont{Klapwijk}}, \bibinfo{journal}{Phys.\ Rev.\ Lett.}
  \textbf{\bibinfo{volume}{91}}, \bibinfo{pages}{046403} 
  (\bibinfo{year}{2003}).

\bibitem[{\citenamefont{Prus et~al.}(2002)\citenamefont{Prus, Reznikov, Sivan,
  and Pudalov}}]{prus02a}
\bibinfo{author}{\bibfnamefont{O.}~\bibnamefont{Prus}},
  \bibinfo{author}{\bibfnamefont{M.}~\bibnamefont{Reznikov}},
  \bibinfo{author}{\bibfnamefont{U.}~\bibnamefont{Sivan}}, \bibnamefont{and}
  \bibinfo{author}{\bibfnamefont{V.}~\bibnamefont{Pudalov}},
  \bibinfo{journal}{Phys.\ Rev.\ Lett.} \textbf{\bibinfo{volume}{88}},
  \bibinfo{pages}{016801} (\bibinfo{year}{2002}).

\bibitem[{\citenamefont{Senz et~al.}(2000)\citenamefont{Senz, Ihn, Heinzel,
  Ensslin, Dehlinger, Gr{\"{u}}tzmacher, and Gennser}}]{senz00}
\bibinfo{author}{\bibfnamefont{V.}~\bibnamefont{Senz}},
  \bibinfo{author}{\bibfnamefont{T.}~\bibnamefont{Ihn}},
  \bibinfo{author}{\bibfnamefont{T.}~\bibnamefont{Heinzel}},
  \bibinfo{author}{\bibfnamefont{K.}~\bibnamefont{Ensslin}},
  \bibinfo{author}{\bibfnamefont{G.}~\bibnamefont{Dehlinger}},
  \bibinfo{author}{\bibfnamefont{D.}~\bibnamefont{Gr{\"{u}}tzmacher}},
  \bibnamefont{and} \bibinfo{author}{\bibfnamefont{U.}~\bibnamefont{Gennser}},
  \bibinfo{journal}{Phys.\ Rev.\ Lett.} \textbf{\bibinfo{volume}{85}},
  \bibinfo{pages}{4357} (\bibinfo{year}{2000}).
  

\end{thebibliography}

\end{document}